# Reactions dynamics for X + $H_2$ insertion reactions (X=C($^1$D), N($^2$D), O($^1$D), S($^1$D)) with Cayley Propagator ring-polymer molecular dynamics

Wenbin Jiang[a], Yuhao Chen[a], Yongle Li[a,b,*]

*a Department of Physics, International Center of Quantum and Molecular Structures, Shanghai University, Shanghai 200444, China*

*b Institute for Quantum Science and Technology Shanghai University, Shanghai Key Laboratory of High Temperature Superconductors, Shanghai University, Shanghai 200444, China*

(Correspondence author: Yongle Li, Email: yongleli@shu.edu.cn)

**ABSTRACT**: In this work, rate coefficients of four prototypical insertion reactions, X + $H_2$ → H + XH (X=C($^1$D), N($^2$D), O($^1$D), S($^1$D)) and associated isotope reactions are calculated based on ring polymer molecular dynamics (RPMD) with Cayley propagator (Cayley-RPMD). The associated kinetic isotope effects (KIEs) are systematically studied too. The Cayley propagator used in this work increases the stability of numerical integration in RPMD calculations, and also supports a larger evolution time interval, allowing us to reach both high accuracy and efficiency. So, our results do not only provide chemical kinetic data for the title reactions in an extended temperature range, but also consist of experimental results, standard RPMD, and other theoretical methods. The results in this work also reflect that Cayley-RPMD has strong consistency and high reliability in the investigations of chemical dynamics for insertion reactions.

## I. INTRODUCTION

The research of chemical reaction rate coefficients plays an indispensable role in the kinetic dynamics modeling of combustion[1], atmospheric[2], and interstellar chemical reactions[3], and provides specific parameters for modeling. And associated kinetic isotope effects (KIEs), which is the same reaction with different nuclides,[4-6] play an indispensable role in investigating the reaction mechanism since the strongest KIE corresponds to the breaking/forming of a bond including the isotopically substituted atom in each elementary step of a given reaction.[7] KIEs can embody the effects caused by multidimensional quantum effects such as zero-point vibration energy and quantum tunneling in chemical reaction systems. However, the experimental measurements of chemical reaction rate coefficients are often difficult, such as, it's hard to control temperature accurately,[8, 9] and hard to prevent the side reactions. So theoretical calculations are necessary for the research of chemical dynamics, the most popular candidates include quantum dynamics with wave packages (QD),[10, 11] quasi-classical trajectory (QCT)[12-14], and methods based on transition state theory (TST)[15-18]. But for calculations of thermal rate coefficients, the QD methods are hindered by the need for lots of quantum numbers, while the two other faster methods, QCT and TST, would contain large errors from neglecting quantum effects.[19] Now a recently developed alternative, named ring-polymer molecular dynamics (RPMD) is used in obtaining thermal rate coefficients. It's an approximated full-dimensional quantum method and can give highly accurate results within an acceptable calculation time. More recently, our group has adopted a new propagator into rate coefficient calculations with RPMD, Cayley propagator,[20-22] making it even faster and more stable in the chemical dynamic calculations. Recently, our group has successfully applied the Cayley propagator to RPMD (Cayley-RPMD) calculations to obtain bimolecular H-capturing reactions, proving that the Cayley-RPMD method has high efficiency with nearly no loss of accuracy.[23, 24]

In chemical dynamics, X + $H_2$→H + XH (X=C($^1$D), N($^2$D), O($^1$D), S($^1$D)) are prototypes for another important kind of bimolecular reaction, insertion reactions. The calculation of thermal rate coefficients is challenging for such kinds of reactions, since they with low potential energy barrier of barrier-less, exhibiting severe non-Arrhenius features on the Arrhenius plot. Specifically, the rate coefficients are weakly dependent on temperature for low-barrier reactions such as N + $H_2$, which with a barrier of about 2 kcal/mol (1.948 of Zhou PES[25] and 1.80 of Ho PES[26]), or independent of temperature for barrier-less reactions, such as C/O/S + $H_2$. For the low-barrier insertion reactions, the quantum tunneling effect can be significant. For all the insertion reactions, there is a deep potential well corresponding to stable bounded states during the reaction pathway (C + $H_2$: -99.8 kcal/mol, N + $H_2$: -126.3 kcal/mol, O + $H_2$: -168.1 kcal/mol, S + $H_2$: -89.9 kcal/mol).

Due to the unique features described above, the insertion reactions always exhibit complicated quantum and recrossing effects, leaving it difficult to obtain reliable computational results by merely using traditional theoretical calculation methods such as variational TST (VTST), Rice-Ramsperger–Kassel–Marcus (RRKM)[27], QCT, etc. To be specific, the reaction process for low-barrier reactions seriously deviates from one of the fundamental assumptions of the TST method, which is the non-recrossing hypothesis. Based on that hypothesis, the TST method optimizes a dividing surface that divides the reactant region and product region in the phase space, and supposes that the reactive trajectories will never go back to the reactant region once passing the dividing surface.[15-18] Such an assumption is incorrect in the heavily recrossing insertion reactions. QCT propagates classical trajectories with quantum-sampled initial states, and can simulate recrossing, correctly, but cannot simulate zero-point energy (ZPE) well, with the well-known ZPE leakage.[10-14] Even worse, QCT cannot simulate the quantum tunneling effect which is essential in N + $H_2$. On the other hand, accurate quantum dynamics (QD) with wave packets are computationally expensive for such systems, since lots of rotational and vibrational states must be considered for obtaining thermal rate coefficients.[10, 28] In O + $H_2$, the authors found that the maximum rotational quantum number of the reactant which can reach $j$=99.[29] After all, till there is only a limited number of thermal rate coefficients and KIE from QD for the title reactions available. So in this work, we want to systematically calculate the rate coefficients of the title reactions in an extended temperature range, and obtain associate KIE, to provide more affirmatory evidence for their kinetics.

Besides, this work is also a further exemplar of using Cayley-RPMD for calculating rate coefficients. The results of this work show that KIEs calculated by Cayley-RPMD are also full of precision for title insertion reactions and compared with experimental values, and other theoretical methods. Finally, this work proves that in the calculations of insertion reactions, Cayley-RPMD is also reliable.

## II. METHOD

### a. RPMD

The details of the RPMD theoretical approach are summarized and described elsewhere,[11, 19, 30] and we present only a brief description of the key steps in this work. For the title reactions, the quantum Hamiltonian is as follows:

$$\hat{H} = \hat{T} + \hat{V} = \sum_{i=1}^{3} \frac{|\hat{\boldsymbol{p}}_i|^2}{2m_i} + V(\hat{\boldsymbol{q}}_1, \hat{\boldsymbol{q}}_2, \hat{\boldsymbol{q}}_3), \tag{1}$$

where $\hat{\boldsymbol{p}}_i$, $\hat{\boldsymbol{q}}_i$ and $m_i$ on behalf of the momentum operators, position operators, and mass of the $i^{\text{th}}$ atom, respectively. Using the isomorphism between a quantum system and the corresponding classical ring-polymer system, the Hamiltonian of RPMD is as follows:

$$H_n(\boldsymbol{p}, \boldsymbol{q}) = \sum_{i=1}^{3} \sum_{j=1}^{n} \left( \frac{\left|\boldsymbol{p}_i^{(j)}\right|^2}{2m_i} + \frac{1}{2} m_i \omega_n^2 \left|\boldsymbol{q}_i^{(j)} - \boldsymbol{q}_i^{(j-1)}\right|^2 \right) + \sum_{j=1}^{n} V\left(\boldsymbol{q}_1^{(j)}, \boldsymbol{q}_2^{(j)}, \boldsymbol{q}_3^{(j)}\right), \tag{2}$$

Here, each atom is seen as a ring of *n* beads connected end to end, each bead is connected by harmonic potentials. Depending on the boundary conditions $\boldsymbol{q}_i^{(0)} = \boldsymbol{q}_i^{(n)}$, the frequency of the bead-bead harmonic potential is $\omega_n = (\beta_n \hbar)^{-1}$, of which $\beta_n = \beta / n = (n k_B T)^{-1}$, where $k_B$ is Boltzmann constant. RPMD needs a pair of dividing surfaces for constructing reaction coordinates. The first dividing surface is defined as $R_\infty$ which is infinite distance (in practice a large enough distance) between the centers of mass (COM) and both two reactants (X=C, N, O and S, and $H_2$) (denoted as $\bar{\boldsymbol{R}}$), making the interaction between them negligible. The first dividing surface is placed near the reactant asymptote:[30]

$$s_0(\boldsymbol{q}) = R_\infty - \bar{\boldsymbol{R}}, \tag{3}$$

where $R_\infty$ means that the reactants are far enough apart that the interaction is almost zero. $\bar{\boldsymbol{q}}_i = \frac{1}{n}\sum_{j=1}^{n} \boldsymbol{q}_{i,j}$ is the bead-averaged position.

The second dividing surface is defined with a reference configuration, and it does not affect the results from RPMD calculations:[30, 31]

$$s_1(\boldsymbol{q}) = \max[(|\bar{\boldsymbol{r}}_{XH_i}| - r_{XH_i}^{\ddagger}) - (|\bar{\boldsymbol{r}}_{HH_i}| - r_{HH_i}^{\ddagger})], \tag{4}$$

where $\bar{\boldsymbol{r}}_{AB}$ is the vector between the centroid of atoms A and B, and $\bar{\boldsymbol{r}}_{AB}^{\ddagger}$ is the distance between atoms A and B at the reference configuration. In general, the reference configuration is the transition state, but in the case of barrier-less reactions, it can be chosen arbitrarily[32, 33]. In this work, the reference configuration for $s_1$ near the bottom of the well, and the detail of the configurations can be found in Table S1 in SI. The reaction coordinates $\xi$ then can be expressed according to the pair of dividing surfaces:[30]

$$\xi(\bar{\boldsymbol{q}}) = \frac{s_0(\bar{\boldsymbol{q}})}{s_0(\bar{\boldsymbol{q}}) - s_1(\bar{\boldsymbol{q}})}, \tag{5}$$

The RPMD rate coefficient is expressed as the multiplication of the following parts using the Bennett-Chandler factorization:[34]

$$k_{\text{RPMD}} = k_{\text{QTST}}(T;\xi^{\ddagger})\kappa(t \to \infty;\xi^{\ddagger}) f(T), \tag{6}$$

Where the $k_{\text{QTST}}(T;\xi^{\ddagger})$ is the static contribution.[35] In practice, $k_{\text{QTST}}(T;\xi^{\ddagger})$ calculated from the peak position $\xi^{\ddagger}$ of the centroid potential of the mean force (PMF) curve along the reaction coordinate $\xi(\boldsymbol{r}_{\text{AB}})$.

$$k_{\text{QTST}}(T;\xi^{\ddagger}) = \frac{4\pi R_{\infty}^2}{\sqrt{2\pi\beta\mu}} e^{-\beta[W(\xi^{\ddagger})-W(0)]}, \tag{7}$$

Here, $\mu$ is the reduced mass of reactants. The $W(\xi^{\ddagger}) - W(0)$ represents the height of the potential of the mean force along the reaction coordinate defined above and is usually obtained by umbrella integration.[36, 37]

The second term $\kappa(t \to \infty;\xi^{\ddagger})$ is named as transmission coefficient and reflects the dynamic contribution of the rate coefficient. It is defined as the ratio of the long-time limit and the zero-time limit of the flux-side correlation function:

$$\kappa(t \to \infty;\xi^{\ddagger}) = \frac{c_{fs}^{(n)}(t \to \infty;\xi^{\ddagger})}{c_{fs}^{(n)}(t \to 0_{+};\xi^{\ddagger})}, \tag{8}$$

The third term $f(T)$ is the contribution of the degeneracy of electronic states. which is equal to the ratio of the electron partition function of the transition state (TS) to the reactant.

$$f(T) = \frac{Q_{\text{elec}}^{\text{TS}}}{Q_{\text{elec}}^{\text{Reactant}}} = \frac{1}{5}, \tag{9}$$

The number of beads has a large impact on the accuracy of the calculated rate, and using single-bead simulations, the RPMD rate is close to the classical TST rate. With the increased number of beads, both accuracy and simulation time increase, with quantum effects simulated more accurately. The minimum number of beads can be calculated using the following formula:[38]

$$n_{\min} = \beta\hbar\omega_{\max}, \tag{10}$$

Here $\omega_{\max}$ is the largest vibrational frequency of the reaction system.

**b. Cayley Propagator**

The details for applying the Cayley operator into RPMD were described in our previous work ($OH + CH_4$),[23] here we only give a summary. The propagator of standard RPMD evolves the coordinates and momenta of the ring-polymer according to the harmonic normal modes in the PIMD method,[37, 39, 40] and can be expressed as below:

$$\begin{bmatrix} \boldsymbol{q}_i(t+\Delta t) \\ \boldsymbol{v}_i(t+\Delta t) \end{bmatrix} = e^{\boldsymbol{L}_i \Delta t} \begin{bmatrix} \boldsymbol{q}_i(t) \\ \boldsymbol{v}_i(t) \end{bmatrix}, \tag{11}$$

Here the $\boldsymbol{q}_i$ and $\boldsymbol{v}_i$ are the position and velocity operators of the $i^{\text{th}}$ atom separately.

Using the Trotter factorization, the $L_i$ is time-evolution operator can be expressed in the following form:

$$e^{\boldsymbol{B}\Delta t/2} e^{\boldsymbol{A}\Delta t/2} e^{\boldsymbol{O}\Delta t} e^{\boldsymbol{A}\Delta t/2} e^{\boldsymbol{B}\Delta t/2} \tag{12}$$

This integrator scheme is called the "BCOCB" scheme,[22] and the operators $\boldsymbol{A}$, $\boldsymbol{B}$ and $\boldsymbol{O}$ include the purely harmonic free ring-polymer motion ( $\boldsymbol{A}$ ), the external potential gradient ( $\boldsymbol{B}$ ), and a thermostat ( $\boldsymbol{O}$ )[41, 42]. Replacing $\boldsymbol{A}$ in the integrator with Cayley propagators, which is written in the following form:[43]

$$\exp(\Delta t\boldsymbol{A}/2) \rightarrow [\text{cay}(\Delta t\boldsymbol{A})]^{1/2}, \tag{13}$$

Here the Cayley propagation matrix can be written as:

$$\text{cay}(\Delta t\boldsymbol{A})^{1/2} = \frac{1}{\sqrt{4+\omega_n^2\Delta t^2}} \begin{pmatrix} 2 & \Delta t \\ -\omega_n^2\Delta t & 2 \end{pmatrix} \tag{14}$$

## c. Computational details

PESs used in this work are nearly the same as in our previous work.[33] In this work, another PES for N + $H_2$ reaction proposed by Zhou *et al.* is also adopted for comparison, which is fitted to more than 20,000 *ab initio* points.[25]

RPMD with Cayley propagator was used to calculate C + $H_2$, N + $H_2$, O + $H_2$, and S + $H_2$ reactions, respectively. For C + $H_2$ we used 32 beads to calculate the rate coefficients at 400 K, 300 K, and 200 K, and we used 64 beads at 127 K. For N + $H_2$, the rate coefficients at 400 K, 300 K, 270 K, and 243 K were calculated with 32 beads. For the O + $H_2$ reaction, the temperatures used at 420 K, 300 K, and 204 K with 32 beads. For S + $H_2$, we calculated at 420 K, 300 K, and 204 K with 128 beads. The time step of both transmission coefficient and free energy is 0.5 fs, which is 5 times larger than that of traditional RPMD, and Cayley-RPMD only took one-fifth of the time of RPMD. Results from this test are also included in the Supporting Information**.** The program used in this work is a customized version of RPMDrate package[30]. The parameters for Cayley-RPMD rate calculations, efficiency, and convergence validation in SI.

# III. RESULTS AND DISCUSSION

## a. X + $H_2$ → H + XH (X=C($^1$D), N($^2$D), O($^1$D), S($^1$D))

**Table 1. Comparison of Cayley-RPMD Rate Coefficients of the X + $H_2$ Reactions with Experimental Values and Results Calculated by Other Theoretical Methods.** The rate

coefficients are in $cm^3 \cdot molecules^{-1} \cdot s^{-1}$, $\Delta G$ is in kcal/mol, the temperatures are in Kelvin, $\xi^{\ddagger}$ and $\kappa(t\rightarrow\infty)$ are dimensionless.

| $C(^1D) + H_2$ | | | | |
|---|---|---|---|---|
| $T$ | 127 | 200 | 300 | 400 |
| $N_{beads}$ | 64 | 32 | 32 | 32 |
| $\xi^{\ddagger}$ | 0.512 | 0.532 | 0.636 | 0.760 |
| $\Delta G(\xi^{\ddagger})$ | 0.210 | 0.346 | 0.510 | 0.760 |
| $k_{QTST}$ | $1.94\times10^{-9}$ | $2.35\times10^{-9}$ | $2.86\times10^{-9}$ | $2.99\times10^{-9}$ |
| $\kappa(t\rightarrow\infty)$ | 0.478 | 0.410 | 0.342 | 0.318 |
| $k_{RPMD}$ | $1.85\times10^{-10}$ | $1.93\times10^{-10}$ | $1.95\times10^{-10}$ | $1.90\times10^{-10}$ |
| $k_{RPMD}^{st,1}$ [33] | … | $1.96\times10^{-10}$ | $1.92\times10^{-10}$ | $1.93\times10^{-10}$ |
| $k_{RPMD}^{st,2}$ [33] | … | $1.86\times10^{-10}$ | $1.93\times10^{-10}$ | $1.92\times10^{-10}$ |
| QD[44] | … | … | $1.50\times10^{-10}$ | … |
| Expt[45] | … | … | $(2.0\pm0.6)\times10^{-10}$ | … |
| $N(^2D) + H_2$（PES of Ho *et al*） | | | | |
| $T$ | 243 | 270 | 300 | 400 |
| $N_{beads}$ | 32 | 32 | 32 | 32 |
| $\xi^{\ddagger}$ | 1.003 | 1.002 | 1.015 | 1.015 |
| $\Delta G(\xi^{\ddagger})$ | 2.305 | 2.397 | 2.697 | 3.180 |
| $k_{QTST}$ | $5.11\times10^{-11}$ | $7.22\times10^{-11}$ | $7.19\times10^{-11}$ | $1.41\times10^{-10}$ |
| $\kappa(t\rightarrow\infty)$ | 0.167 | 0.161 | 0.224 | 0.252 |
| $k_{RPMD}$ | $1.71\times10^{-12}$ | $2.34\times10^{-12}$ | $3.22\times10^{-12}$ | $7.10\times10^{-12}$ |
| $k_{RPMD}^{st,1}$ [33] | … | $2.18\times10^{-12}$ | $3.15\times10^{-12}$ | $7.04\times10^{-12}$ |
| $k_{RPMD}^{st,2}$ [33] | … | $2.23\times10^{-12}$ | $3.26\times10^{-12}$ | $6.79\times10^{-12}$ |
| QD[46] | … | $2.05\times10^{-12}$ | $2.87\times10^{-12}$ | $6.36\times10^{-12}$ |
| Expt[47] | $(1.67\pm0.15)\times10^{-12}$ | $(1.67\pm0.15)\times10^{-12}$ | $(2.44\pm0.34)\times10^{-12}$ | … |
| $N(^2D) + H_2$（PES of Zhou *et al*） | | | | |
| $T$ | 243 | 270 | 300 | 400 |
| $N_{beads}$ | 32 | 32 | 32 | 32 |
| $\xi^{\ddagger}$ | 1.000 | 1.010 | 1.013 | 1.017 |
| $\Delta G(\xi^{\ddagger})$ | 2.512 | 2.651 | 2.850 | 3.400 |
| $k_{QTST}$ | $2.98\times10^{-11}$ | $4.03\times10^{-11}$ | $5.64\times10^{-11}$ | $1.07\times10^{-10}$ |
| $\kappa(t\rightarrow\infty)$ | 0.204 | 0.210 | 0.221 | 0.256 |
| $k_{RPMD}$ | $1.22\times10^{-12}$ | $1.70\times10^{-12}$ | $2.50\times10^{-12}$ | $5.50\times10^{-12}$ |
| $k_{RPMD}^{st}$ [32] | … | … | $3.21\times10^{-12}$ | $6.98\times10^{-12}$ |
| QD[48] | … | $1.72\times10^{-12}$ | $2.44\times10^{-12}$ | $5.63\times10^{-12}$ |
| Expt[47] | $(1.39\pm0.06)\times10^{-12}$ | $(1.67\pm0.15)\times10^{-12}$ | $(2.44\pm0.34)\times10^{-12}$ | … |
| $O(^1D) + H_2$ | | | | |
| $T$ | … | 204 | 300 | 420 |
| $N_{beads}$ | … | 32 | 32 | 32 |
| $\xi^{\ddagger}$ | … | 0.849 | 0.860 | 0.868 |
| $\Delta G(\xi^{\ddagger})$ | … | 0.853 | 0.830 | 1.337 |
| $k_{QTST}$ | … | $1.81\times10^{-9}$ | $1.66\times10^{-9}$ | $1.60\times10^{-9}$ |
| $\kappa(t\rightarrow\infty)$ | … | 0.305 | 0.341 | 0.361 |
| $k_{RPMD}$ | … | $1.10\times10^{-10}$ | $1.13\times10^{-10}$ | $1.18\times10^{-10}$ |
| $k_{RPMD}^{st,1}$ [33] | … | … | $1.12\times10^{-10}$ | $1.17\times10^{-10}$ |
| $k_{RPMD}^{st,2}$ [33] | … | … | $1.13\times10^{-10}$ | $1.16\times10^{-10}$ |
| $k_{RPMD}^{st}$ [32] | … | … | $1.11\times10^{-10}$ | $1.14\times10^{-10}$ |
| QD1[49] | … | … | $9.90\times10^{-11}$ | $1.01\times10^{-10}$ |
| QD2[50] | … | … | $1.00\times10^{-10}$ | … |
| QD3[51] | … | … | $1.25\times10^{-10}$ | $1.25\times10^{-10}$ |
| Expt[52] | … | … | $(1.20\pm0.18)\times10^{-10}$ | … |
| Expt[53] | … | … | $(1.03\pm0.04)\times10^{-10}$ | … |
| Expt[48] | … | $1.0\times10^{-10}$ | | … |
| $S(^1D) + H_2$ | | | | |
| $T$ | … | 200 | 300 | 400 |
| $N_{beads}$ | … | 128 | 128 | 128 |
| $\xi^{\ddagger}$ | … | 0.677 | 0.916 | 0.922 |
| $\Delta G(\xi^{\ddagger})$ | … | 0.231 | 0.507 | 0.853 |
| $k_{QTST}$ | … | $2.85\times10^{-9}$ | $2.76\times10^{-9}$ | $2.57\times10^{-9}$ |
| $\kappa(t\rightarrow\infty)$ | … | 0.285 | 0.294 | 0.317 |
| $k_{RPMD}$ | … | $1.62\times10^{-10}$ | $1.62\times10^{-10}$ | $1.63\times10^{-10}$ |
| $k_{RPMD}^{st,1}$ [33] | … | $1.57\times10^{-10}$ | $1.64\times10^{-10}$ | $1.67\times10^{-10}$ |
| $k_{RPMD}^{st,2}$ [33] | … | $1.59\times10^{-10}$ | $1.65\times10^{-10}$ | $1.65\times10^{-10}$ |
| QD[54] | … | … | $1.49\times10^{-10}$ | … |
| Expt[55] | … | … | $(2.10\pm0.21)\times10^{-10}$ | … |
| Expt[56] | … | … | $(2.07\pm0.21)\times10^{-10}$ | … |

"st" here means standard RPMD rate.

"1" here means the dividing surface is set in front of the energy barrier.

"2" here means the dividing surface is set at the potential well.

The results of the four title reactions calculated by Cayley-RPMD of the number of beads ($N_{\text{beads}}$), the $\xi^{\ddagger}$, associated $\Delta G(\xi^{\ddagger})$, $k_{\text{QTST}}$, $\kappa(t \to \infty)$, and $k_{\text{RPMD}}$ at different temperatures are summarized in Table 1, with the comparison of calculated results from the standard RPMD ($k_{\text{RPMD}}^{\text{st}}$) or other theoretical methods, and experiments. In Table S1 of SI, we have also reported the minimum number of beads for each system at each temperature. The free energy barriers of C/N/O/S + $H_2$ are all very small, only those for N + $H_2$ are larger, as 2.3~3.4 kcal/mol. Since Zhou PES has a higher potential barrier at TS, the $\Delta G(\xi^{\ddagger})$ values at all the temperatures investigated in this work on Zhou PES are also larger than those on Ho PES. The transmission coefficients of RPMD with the Cayley propagator are generally lower than standard RPMD, and the corresponding free energy barriers are also lower, leaving the overall results similar. This feature of results from Cayley-RPMD is similar to those found in our previous works.[23, 24] Finally, the relative deviations of all the calculated $k_{\text{RPMD}}$ from Cayley-RPMD are within 5% from standard RPMD, and are in good agreement with the experimental results, as observed in standard RPMD from our previous works.[32, 33]

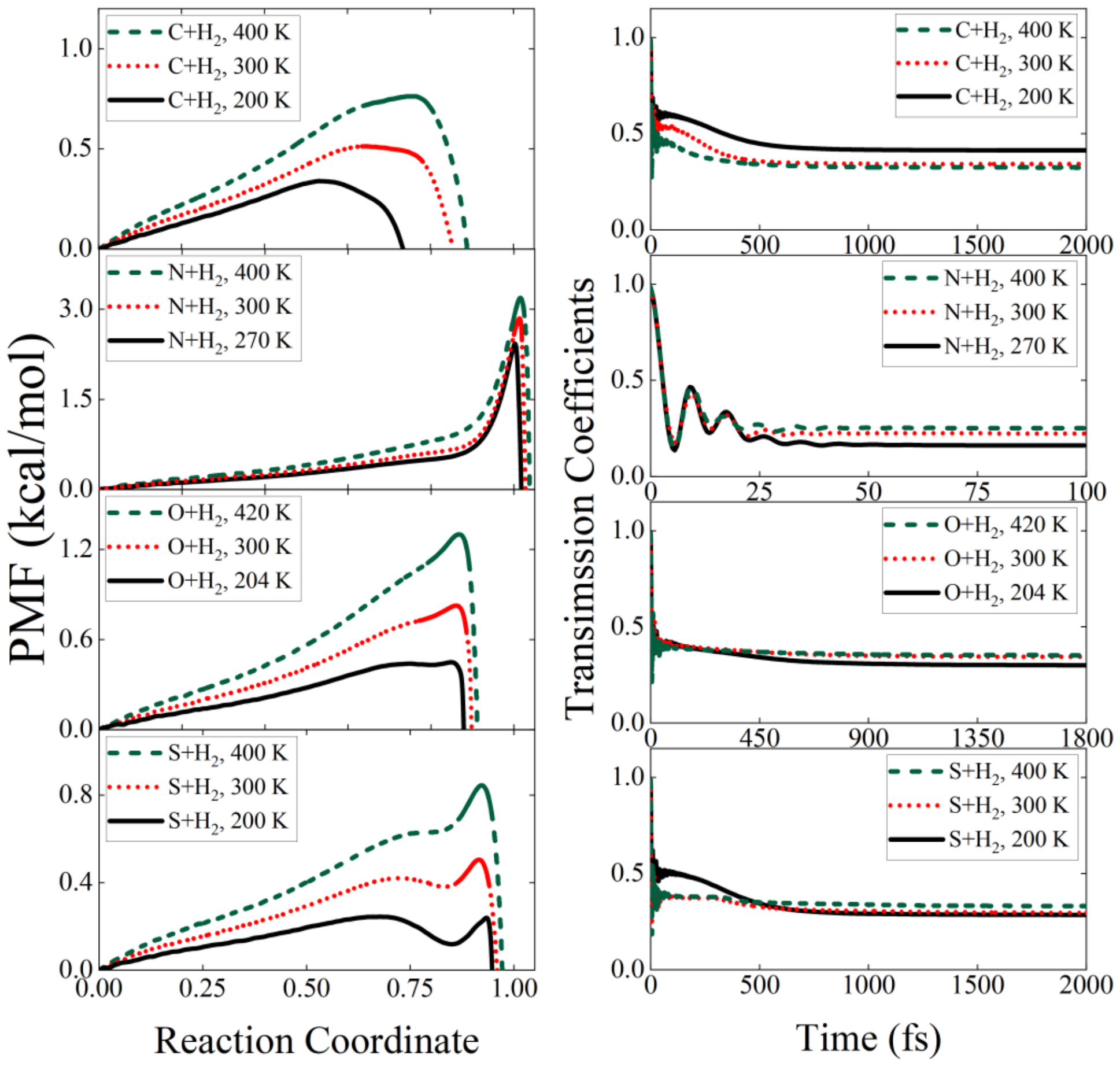

**Figure 1.** Comparison of the potential of mean force (PMFs) (left panel) and transmission coefficient ( $\kappa(t)$ ) (right panel) for C/N/O/S + $H_2$ at different temperatures.

From the curves of $G(\xi)$ in the left panel of Figure 1, it can be seen the details of the peaks of $G(\xi)$ changing with reaction coordinate $\xi$ and temperature: the barrier heights for C/S + $H_2$ are both no more than 1 kcal/mol at all temperatures investigated, those of O + $H_2$ are no more than 1.3 kcal/mol at 420 K, and drop to less than 1 kcal/mol at lower temperatures, and only those for N + $H_2$ are slightly higher but still less than 3.4 kcal/mol since it has a peak on PES. For the three barrierless reactions, we can also observe that the temperature dependence of the barrier heights is quite small. From the transmission coefficient ($\kappa(t)$) curves shown in the right panel of Figure 1, one can find that $\kappa(t)$ curves for C/O/S + $H_2$ exceed 1000 fs to converge (1500 fs for C + $H_2$, 1100 fs for O + $H_2$, and 1500 fs for S + $H_2$) and that for N + $H_2$ needs over 75 fs to reach convergence. The convergence time for insertion reactions is longer for other hydrogen-abstraction reactions[23, 24, 57, 58], for which systems, the convergence times are all within 100 fs. Because in RPMD calculations for barrier-less reactions, one selects dividing surfaces arbitrarily, and trajectories through the chosen dividing surfaces many times. Especially, since the title reactions are all with a deep potential well along the reaction pathway, the reaction trajectories are trapped in the wells for a long time, which can also result in a longer convergence time. In this situation, the Cayley-RPMD[20] can exhibit its power of propagating enough time within an affordable CPU time and without accuracy loss.

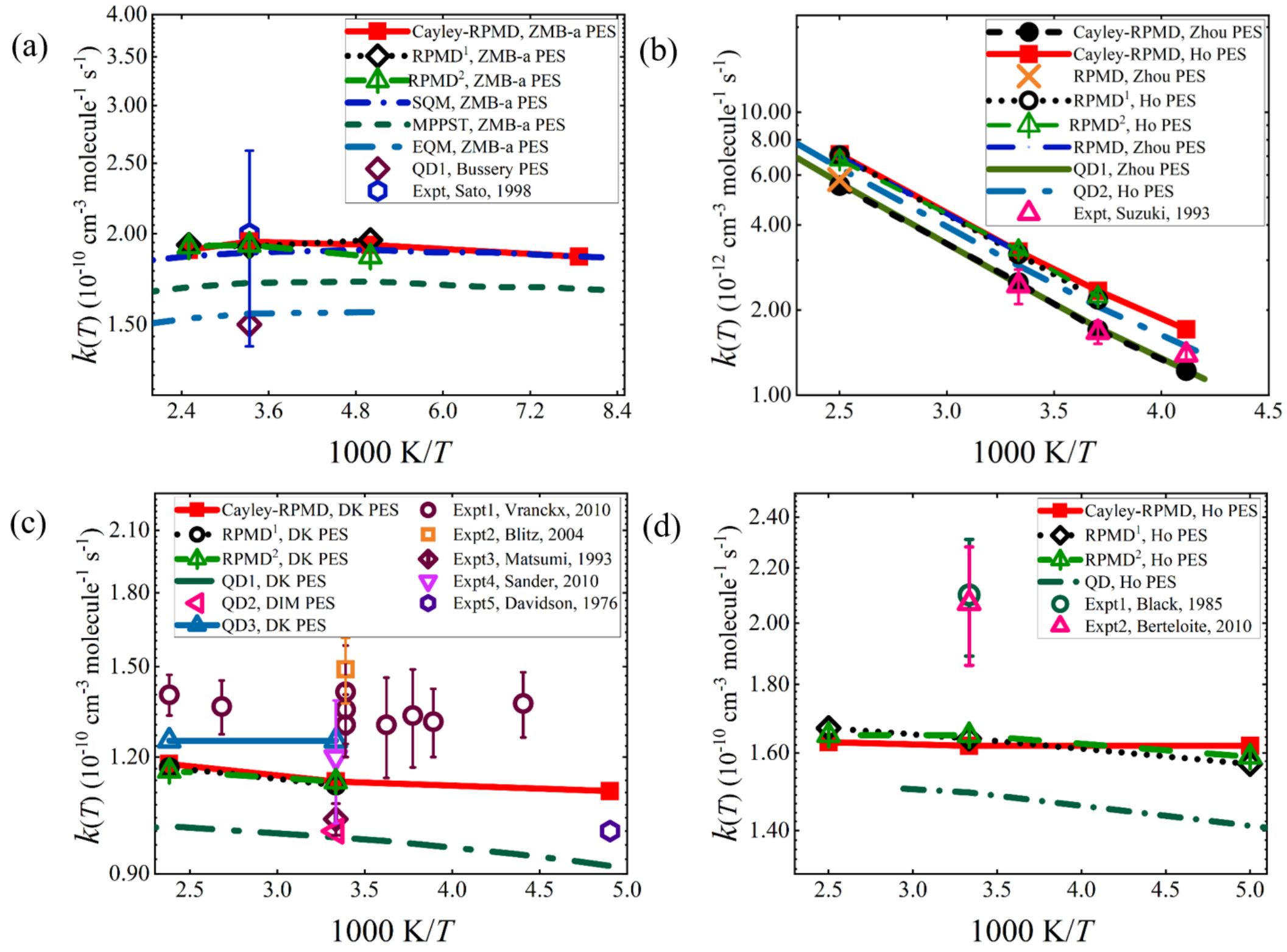

**Figure 2.** Rate coefficients of X(C/N/O/S) + $H_2$ by Cayley-RPMD compared with the experimental values and other theoretical values. ((a), (b), (c), and (d) represent the results of C + $H_2$, N + $H_2$, O + $H_2$, and S + $H_2$ respectively. The rate coefficients of standard RPMD (shown as $RPMD^1$(the dividing surface is set in front of the energy barrier), $RPMD^2$ (the dividing surface is set at the potential well), and RPMD  Zhou PES means result calculated on Zhou PES using the same parameters as Cayley-RPMD in this work.) in the plots, and the same denotation is used in all four subplots) from Wang *et al.*[33]. (a): Cayley-RPMD and standard RPMD results on ZMB-a PES[59], statistical quantum mechanical method (SQM), mean potential phase space theory (MPPST), and exact quantum mechanical method (EQM)[60] results on ZMB-a PES[59], QD[44] on Bussery PES, experimental result[45]. (b): Cayley-RPMD on Zhou-PES[25] and Ho-PES[26], standard RPMD and QD1[46] results on Zhou PES, QD2[48] results on Ho PES, experimental result[47]. (c): Cayley-RPMD and standard RPMD results on DK-PES[48], QD1[49], QD2[50], and QD3[51], experimental results[47, 52, 61-63]. (d): Cayley-RPMD and standard RPMD results on Ho-PES[64], QD[54] results on Ho PES[64], experiments[55, 56].)

Figure 2(a) collects the rate coefficients of C + $H_2$ at 127 K, 200 K, 300 K, and 400 K in this work. There is only a single experimental data at 300 K, $(2.0 \pm 0.6) \times 10^{-10}$ $cm^3 \cdot molecules^{-1} \cdot s^{-1}$. Our result at 300 K is $1.95 \times 10^{-10}$ $cm^3 \cdot molecules^{-1} \cdot s^{-1}$ within the range of experimental error.

The relative deviation of rate coefficients between Cayley-RPMD and the standard RPMD is negligible, within only 4%. The work QD1 uses the quantum reactive scattering method with wave package and on a different PES, so there is a distinguishable deviation between $k(T)$ from QD1 and our work. And results from other theoretical methods, SQM, MPPST, and EQM, in which SQM and MPPST both with assumptions that the reaction takes place through the formation of an intermediate complex between reactants and products. One can observe that the RPMD results are in good accordance with those from SQM, but deviate much from the results from EQM. And those from EQM are close to the QD1 result.

N + $H_2$ reaction is a typical insertion reaction with a low potential barrier as discussed in the section of the Introduction. The $k(T)$ is weekly temperature dependent. Figure 2(b) shows that the N + $H_2$ reaction rate coefficients are calculated by Cayley-RPMD based on Zhou PES and Ho PES, respectively. Results from other theoretical methods and experiments are also collected. The overall temperature dependence of the $k(T)$ for N+$H_2$ is in Arrhenius shape. The relative deviation from Cayley-RPMD and standard RPMD obtained on both two PESs is small, within 5%. It can be found that the Cayley-RPMD obtained on Zhou PES are closer to the results from the experiment and ones from QD, in which the relative deviations are only 1%-6%. This small relative deviation stems from that Zhou PES is more accurate for both potential barrier and shape. And since Zhou PES has a higher barrier than Ho PES (1.948 kcal/mol on Zhou PES vs 1.80 kcal/mol on Ho PES, with a relative deviation of 9%), the results on Zhou PES are smaller. In general, the results from Cayley-RPMD are all close to those from the QD method. The QD1 curve is calculated by H. Guo's group on Zhou PES, using the Chebyshev wave packet method.[46] The Cayley-RPMD results on Zhou PES are close to those from QD1, and within a 2% relative deviation. Since the results from QD1 contain error of approximated rotational quantum effect by using J-shifting, their coincidence to experimental results and results from RPMD is accidental. One can also observe that the new standard RPMD on Zhou PES deviates from our 2014 result[32], but this deviation is within 21%, still within the error range of RPMD. Since in our work of OH + $CH_4$, we have systematically tested the choice of force constant used in umbrella sampling, we have also checked the result from standard RPMD on Zhou PES using the same parameters in Cayley-RPMD this work, at 400 K. That result is $5.78\times10^{-12}$ $cm^3\cdot molecules^{-1}\cdot s^{-1}$ and details are collected in the Supporting Information. This value is with a relative deviation less than 5%, showing the consistency of the Cayley-RPMD with 0.5 fs time interval. For results from the Cayley-RPMD and QD2 method on Ho PES, the deviation is still small, within 13%.

The results of the O + $H_2$ reaction at 420 K, 300 K, and 204 K are collected in Figure 2(c). It can be seen that the results calculated by standard RPMD are similar to the Cayley-RPMD, in which the relative deviations are only 1%. The relative deviation between Cayley-RPMD and QD3 is 11%, QD2 is about 12%, and QD1 is less than 15%. Results from QD1 and QD3 are on DK PES and ones from QD2 are on DIM PES. The difference between the two PES is that the $1A''$ barrier is different (the potential barrier on DIM PES is 3.7 kcal/mol, and on DK PES is 2.3 kcal/mol), which makes the $1A'$ cross-section of DIM PES larger. Results from all the QD methods are from the rotational and vibrational ground state ($v$=$j$=0). Results from both QD1 and QD2 include all three electronic states. In this reaction system, the calculated values of this work are within the range of experimental results. Especially, our results agree well with the experimental values, the deviation between them is within 8% at 300 K and is 10% at 204 K, which is also in line with the common error of standard RPMD, which is reported as 30%. The new result at 204 K confirms further that the reaction rate of O + $H_2$ is independent of temperature at low temperatures.

The reaction rate coefficients from S + $H_2$ at 400 K, 300 K, and 200 K are collected in Figure 2(d). The Cayley-RPMD rate coefficients are in the range of (1.62~1.63) × $10^{-10}$ $cm^3$·$molecules^{-1}$·$s^{-1}$. Again, the rate coefficients are temperature-independent, and results from Cayley-RPMD are even better to show this feature than those from standard RPMD. The different choices of dividing surfaces in RPMD calculations caused the apparent different trends, but actually in number they are quite the same, the maximum relative deviation between the results from Cayley-RPMD and standard RPMD for this reaction is only 4%. Furthermore, since there are only three points, it looks like they are in different trends. Results in our 2019 work[33] of the same choice of dividing surface are more similar with our results in the current work. The 23% difference between the results from Cayley-RPMD and from both experiments would stem from the non-adiabaticity, as observed from our previous work.[65]

### b. X + $D_2$ → D + XD (X=C($^1$D), N($^2$D), O($^1$D), S($^1$D))

**Table 2. Comparison of Cayley-RPMD Rate Coefficients of the X + $D_2$ Reaction with Experimental Values and Results Calculated by Other Theoretical Methods.** The rate coefficients are in $cm^3$·$molecules^{-1}$·$s^{-1}$, $\Delta G$ is in kcal/mol, the temperatures are in Kelvin, $\xi^{\ddagger}$ and $\kappa(t \to \infty)$ are dimensionless.

| C($^1$D) + $D_2$ | | | | |
|---|---|---|---|---|
| $T$ | 127 | 200 | 300 | 400 |
| $N_{\text{beads}}$ | 64 | 32 | 32 | 32 |
| $\xi^{\ddagger}$ | 0.522 | 0.548 | 0.685 | 0.760 |
| $\Delta G(\xi^{\ddagger})$ | 0.220 | 0.369 | 0.553 | 0.830 |
| $k_{\text{QTST}}$ | 1.41×$10^{-9}$ | 1.66×$10^{-9}$ | 2.03×$10^{-9}$ | 2.09×$10^{-9}$ |
| $\kappa(t \to \infty)$ | 0.487 | 0.424 | 0.349 | 0.334 |
| $k_{\text{RPMD}}$ | 1.37×$10^{-10}$ | 1.41 × $10^{-10}$ | 1.42×$10^{-10}$ | 1.40×$10^{-10}$ |
| $k^{\text{st}}_{\text{RPMD}}$ [66] | 9.98×$10^{-11}$ | … | 1.17×$10^{-10}$ | … |
| Expt[45] | … | … | (1.4±0.3)×$10^{-10}$ | … |
| N($^2$D) + $D_2$（PES of Ho *et al*） | | | | |
| $T$ | 243 | 270 | 300 | 400 |
| $N_{\text{beads}}$ | 32 | 32 | 32 | 32 |
| $\xi^{\ddagger}$ | 1.015 | 1.016 | 1.017 | 1.019 |
| $\Delta G(\xi^{\ddagger})$ | 2.582 | 2.720 | 2.881 | 3.365 |
| $k_{\mathrm{QTST}}$ | $2.10\times10^{-11}$ | $2.93\times10^{-11}$ | $3.98\times10^{-11}$ | $8.45\times10^{-11}$ |
| $\kappa(t\to\infty)$ | 0.200 | 0.206 | 0.216 | 0.242 |
| $k_{\mathrm{RPMD}}$ | $8.40\times10^{-13}$ | $1.21\times10^{-12}$ | $1.72\times10^{-12}$ | $4.09\times10^{-12}$ |
| Expt[47] | $(8\pm0.4)\times10^{-13}$ | $(1.12\pm0.04)\times10^{-12}$ | … | … |
| $N(^2D) + D_2$（PES of Zhou *et al*） | | | | |
| $T$ | 243 | 270 | 300 | 400 |
| $N_{\mathrm{beads}}$ | 32 | 32 | 32 | 32 |
| $\xi^{\ddagger}$ | 1.015 | 1.017 | 1.017 | 1.020 |
| $\Delta G(\xi^{\ddagger})$ | 2.858 | 2.581 | 3.089 | 3.688 |
| $k_{\mathrm{QTST}}$ | $1.43\times10^{-11}$ | $1.92\times10^{-11}$ | $2.76\times10^{-11}$ | $5.54\times10^{-11}$ |
| $\kappa(t\to\infty)$ | 0.199 | 0.218 | 0.224 | 0.255 |
| $k_{\mathrm{RPMD}}$ | $5.70\times10^{-13}$ | $8.26\times10^{-13}$ | $1.24\times10^{-12}$ | $2.82\times10^{-12}$ |
| Expt1[47] | $(8\pm0.4)\times10^{-13}$ | $(1.12\pm0.04)\times10^{-12}$ | … | … |
| Expt2[67] | … | … | $(1.37\pm0.19)\times10^{-12}$ | … |
| Expt3[68] | … | … | $(1.60\pm0.34)\times10^{-12}$ | … |
| $O(^1D) + D_2$ | | | | |
| $T$ | … | 204 | 300 | 420 |
| $N_{\mathrm{beads}}$ | … | 32 | 32 | 32 |
| $\xi^{\ddagger}$ | … | 0.722 | 0.865 | 0.870 |
| $\Delta G(\xi^{\ddagger})$ | … | 0.461 | 0.830 | 0.876 |
| $k_{\mathrm{QTST}}$ | … | $1.33\times10^{-9}$ | $1.93\times10^{-9}$ | $1.18\times10^{-9}$ |
| $\kappa(t\to\infty)$ | … | 0.309 | 0.345 | 0.373 |
| $k_{\mathrm{RPMD}}$ | … | $8.22\times10^{-11}$ | $8.40\times10^{-11}$ | $8.80\times10^{-11}$ |
| QD[69] | … | … | $7.6\times10^{-11}$ | $1.01\times10^{-10}$ |
| QD[69] | … | … | $1.30\times10^{-10}$ | … |
| Expt1[61] | … | … | $(1.30\pm0.05)\times10^{-10}$ | … |
| Expt2[70] | … | … | $1.00\times10^{-10}$ | … |
| $S(^1D) + D_2$ | | | | |
| $T$ | … | 200 | 300 | 400 |
| $N_{\mathrm{beads}}$ | … | 128 | 128 | 128 |
| $\xi^{\ddagger}$ | … | 0.667 | 0.920 | 0.924 |
| $\Delta G(\xi^{\ddagger})$ | … | 0.254 | 0.530 | 0.876 |
| $k_{\mathrm{QTST}}$ | … | $2.02\times10^{-9}$ | $1.93\times10^{-9}$ | $1.79\times10^{-9}$ |
| $\kappa(t \to \infty)$ | … | 0.281 | 0.298 | 0.322 |
| $k_{RPMD}$ | … | $1.14\times10^{-10}$ | $1.15\times10^{-10}$ | $1.16\times10^{-10}$ |

Table 2. shows the reaction rates for the isotopic reactions for the title reactions, X + $D_2$ → D + XD (X=C($^1$D), N($^2$D), O($^1$D), S($^1$D)) in different temperatures by Cayley-RPMD and other theoretical methods. Again, the results are in good accordance with all the experimental results.

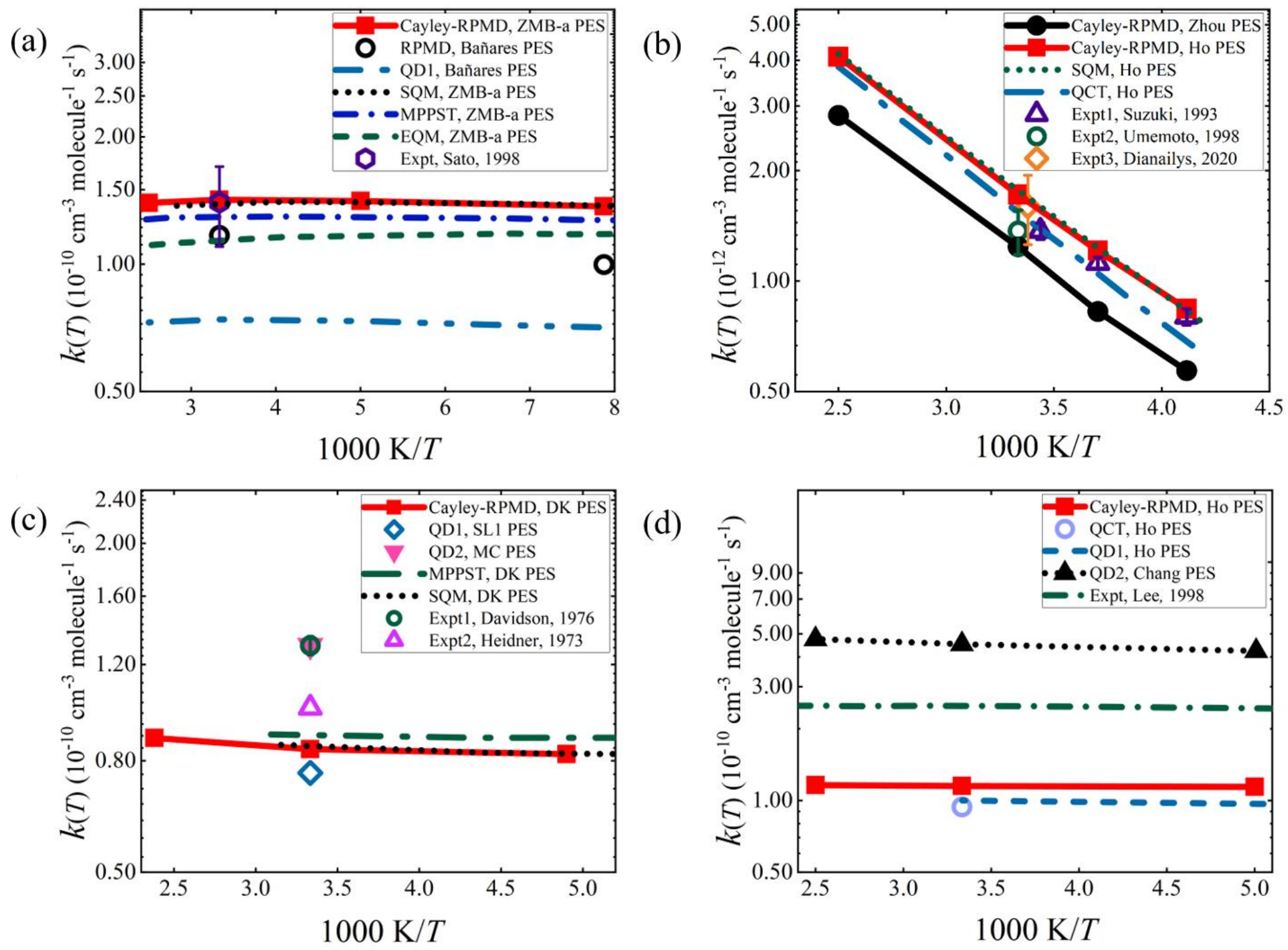

**Figure 3.** Comparison of rate coefficients by Cayley-RPMD and the experimental values, and theoretical values. ((a), (b), (c), and (d) represent the results of C/N/O/S + $D_2$, respectively.) (a): Cayley-RPMD on ZMB-a PES[59], standard RPMD[66] results on Bañares PES[71], QD1 value from Lin *et al.*[72] on Bañares PES, Statistical quantum mechanical method (SQM), Mean potential phase space theory (MPPST), and Exact quantum mechanical method (EQM) are Lezana *et al.*'s[60] results on ZMB-a PES, experimental result[45]. (b): Cayley-RPMD on Zhou-PES[25] and Ho-PES[26], SQM[73] and QCT[73] results on Ho-PES, experimental results[47, 67, 68]. (c): Cayley-RPMD results on DK-PES[48], QD1[69] on SL1 PES, QD2[69] on MC PES, SQM[70] and MPPST[70] on DK-PES, experimental results[61, 70]. (d): Cayley-RPMD and standard RPMD results on Ho PES[64], QCT on Ho PES[54], QD1[74] on Ho-PES, QD2[75] on Chang PES, experimental result[76].)

For C + $D_2$ from Cayley-RPMD on ZMB-a PES at 127 K, 200 K, 300 K, and 400 K are collected in Figure 3(a). At 300 K, the result from Cayley-RPMD has a relative deviation from the experimental value of less than 2%. SQM results on ZMB-a PES are nearly the same as those from Cayley-RPMD, while MPPST results deviate slightly, with an average deviation of about 3%. The EQM results again deviate from those from Cayley-RPMD. The relative deviation is 25% between standard RPMD from Hickson *et al.*[66] and results in this work, mainly from the difference between PESs. The ZMB-a PES is more accurate than the Bañares PES. Results from QD1 are obtained by Lin *et al.*[72] using the Chebyshev wave packet method on Bañares PES, with the relative deviation to results from both Cayley-RPMD and experiments being about 50%. This deviation would come from the insufficient quantum number used in their work, which only includes the rotational and vibrational ground state ($v$=$j$=0).

Figure 3(b) collects the results of N + $D_2$ reaction based on Ho PES and Zhou PES, respectively. It also contains results from SQM and QCT on Ho PES.[73] Expt1 from Suzuki *et al.*[47], Expt2 is from Umemoto *et al.*[67], and Expt3 is from Dianailys *et al.*[68] Again, the relative deviation between results from Cayley-RPMD and ones from SQM on Ho PES is very small, within 3%. The relative deviation between results from Cayley-RPMD and ones from QCT is within 20% on Ho PES and about 35% on Zhou PES. This deviation stems from the combined effects of both ZPE leakage and ignoring quantum tunneling in QCT. Since the barrier on Ho PES is lower than that on Zhou PES, the calculated results are smaller on Zhou PES. Interestingly, the results on Zhou PES form the lower bound and those on Ho PES form the upper bound.

The results of the O + $D_2$ reaction on DK PES[48] are shown in Figure 3(c). The relative deviation between results from this work and those from SQM is about only 1%, and that between results from this work and those from MPPST is less than 3%. QD1 and QD2 are both from adiabatic capture/infinite order sudden approximation (ACIOSA) model[69], but on different PESs. The deviation between results from Cayley-RPMD and that from Expt2 is less than 20%, but that between results from Cayley-RPMD and ones from Expt1 is as large as 36%. The large deviation would stem from the experimental overestimation, since in Expt1, a buffer gas is used, and such finite pressure can cause the rate increase.

Figure 3(d) shows the rate coefficients of the S($^1$D) + $D_2$ reaction on Ho PES. The relative deviation between results in this work and ones from QD1 is about 15%, and the deviation to results from QCT is about 18%, which is due to the QCT ignoring the zero-point energy and tunneling effect of the reaction. The results of S($^1$D) + $H_2$ and S($^1$D) + $D_2$ calculated by QD2 are 5-6 times our results. QD2 is calculated by variable RRKM theory based on another *ab initio* PES. The experimental results are obtained by fitting the activation function with parameters using data measured from the cross-molecular beam experiment. The relative deviation between the results from this work and the experiment is larger, at about 50%. This deviation would come from the nonadiabatic effect among different electronic states along the reaction path.

**c. KIE**

**Table 3. Comparison of KIEs of the Title Reactions with Experimental Values and Results Calculated by Other Theoretical Methods.**

| C($^1$D) | | | | |
|---|---|---|---|---|
| $T$/ K | 127 | 200 | 300 | 400 |
| Cayley-RPMD | 1.35 | 1.37 | 1.37 | 1.36 |
| RPMD[st 66] | … | … | 1.37 | … |
| Expt[45] | … | … | 1.43 | … |
| N($^2$D) | | | | |
| *T*/ K | 243 | 270 | 300 | 400 |
| Cayley-RPMD (Ho PES) | 2.04 | 1.93 | 1.87 | 1.74 |
| Cayley-RPMD (Zhou PES) | 2.14 | 2.06 | 2.02 | 1.94 |
| QD1[48] | … | … | 1.76 | 1.64 |
| Expt[47] | 1.74 | 1.49 | … | … |
| O($^1$D) | | | | |
| *T*/ K | … | 204 | 300 | 420 |
| Cayley-RPMD | … | 1.34 | 1.35 | 1.34 |
| QD[69] | … | … | 1.33 | … |
| QD[69] | … | … | 1.35 | … |
| Expt[45] | … | … | 1.00 | … |
| Expt[77] | … | … | 1.50 | … |
| S($^1$D) | | | | |
| *T*/ K | … | 200 | 300 | 400 |
| Cayley-RPMD | … | 1.42 | 1.41 | 1.41 |
| QD[75] | … | 1.37 | 1.37 | 1.37 |

*"st"* means KIEs from classical RPMD rate.

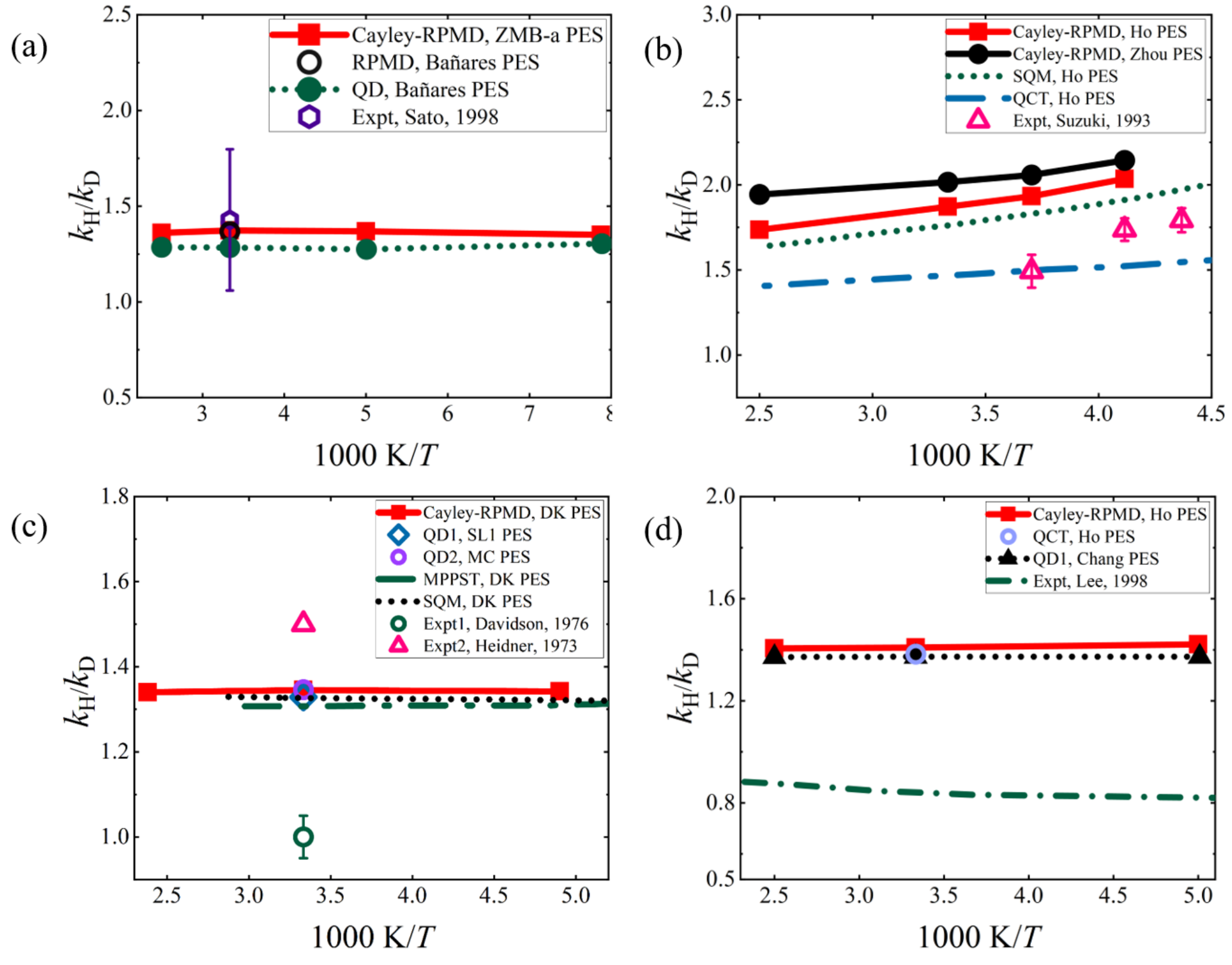

**Figure 4.** Comparison of KIEs ( $k_H / k_D$ ) by RPMD (Cayley) and the experimental values、theoretical values. ((a), (b), (c), and (d) represent the results of C/N/O/S, respectively. (a): Cayley-RPMD on ZMB-a PES[59], standard RPMD[66, 78] results on Bañares PES, QD values from Lin *et al.*[72] on Bañares PES, experimental result[45]. (b): Cayley-RPMD on Zhou-PES[25] and Ho-PES[26], SQM[73] and QCT[73] results on Ho-PES, experimental result[47]. (c): Cayley-RPMD results on DK-PES[48], QD1[69] on SL1 PES, QD2[69] on MC PES, SQM[77] and MPPST[77] on DK-PES, experimental results[45, 69]. (d): Cayley-RPMD and standard RPMD results on Ho PES[64], QCT[74] results on Ho-PES, QD1[75] on Chang PES, experimental result[76])

Figure 4(a) collects the KIE of C($^1$D) + $H_2/D_2$. The relative deviation between results from Cayley-RPMD and those from experiments of Sato *et al.*[45] is only about 4%. The relative deviation between the Cayley-RPMD and all the theoretical values is very small, less than 8%. The values are different but the relative deviation of KIE results between Cayley-RPMD and standard RPMD is less than 1%. QD values are Lin *et al.*[72] by wave packet method. The relative deviation compared with QD is less than 8%, which is from the inaccuracy of PES.

The KIE of N($^2$D) + $H_2/D_2$ are shown in Figure 4(b). The relative deviation of KIEs result between Cayley-RPMD and SQM on Ho PES is less than 10%. The relative deviation of the Cayley-RPMD results between ones obtained from Zhou PES and Ho PES is about 15%. It shows that the difference in PES affects much to KIE's result. The results from QCT change less with temperature, but those from Cayley-RPMD and SQM increase with lowering temperature. From our knowledge and discussion in the previous work on QCT and SQM, this comes from the inclusion of tunneling in RPMD and SQM and agrees more with the experimental results. At 243 K, the deviation between the experimental value and results from Cayley-RPMD on Zhou PES is 24%, and those on Ho PES is 18%. But when the temperature goes to 270 K, the deviations between results from Cayley-RPMD on both PESs and the experimental value increase to 40%, as well as that from SQM. Maybe it stems from the non-adiabatic coupling of electronic states since the shape of Zhou PES is quite good.

Figure 4(c) collects $O(^1D) + H_2/D_2$ of KIE on DK PES. The KIEs' relative deviation is smaller than that of SQM and MPPST, which reflects the determining factor of KIEs is the potential energy surface. The error of Expt2 is less than 11%, and the error between Expt1 and our result is about 35%. All the theoretical calculated values are similar and fall within the two experimental ranges. The larger error may be due to the unbalanced $H_2$ mixture used in the experiment.

$S(^1D) + H_2/D_2$ of KIE by Cayley-RPMD are shown in Figure 4(d). For this reaction, the experimental results are fitted from crossed molecular beam[76]. And all the theoretical results are nearly the same, even on different PESs. However, all the theoretical results deviate from the experimental ones by about 70%. Since again this reaction has several electronic states, further non-adiabatic calculations would give more convincing results.

In summary, we found that in the insertion reactions, the rate coefficients from Cayley-RPMD are highly compatible to the experimental results and results from other theoretical methods. The results from Cayley-RPMD are within deviation of 23% for $X + H_2$ ($C/N/O + H_2$ within deviation of 10%, $S + H_2$ within deviation of 23%) and 50% for $X + D_2$ ($C/N/O + D_2$ within deviation of 20%, $S + D_2$ within deviation of 50%), to the experimental results. And in this work, the deviation between results from Cayley-RPMD and standard RPMD is also systematically compared. We found with the help of the highly stability of Cayley propagator, we can achieve a much larger time interval of simulation as 0.5 fs, with deviation within 5%. This is cheering since the enhancement of efficiency of RPMD calculations allows us to extend the scale of the calculations of rate coefficients in more complicated reactions.

## IV. Conclusions

In this work, the $X + H_2 \rightarrow H + XH$ ($X=C(^1D), N(^2D), O(^1D), S(^1D)$) reactions and associated KIE ($X + H_2/D_2$) are calculated by highly stable and efficient Cayley-RPMD in a larger temperature range comparing with our previous work[33]. The overall relative deviation among all the results is less than 5% compared with the standard RPMD. With more results, we have confirmed the temperature-irrelevant feature for all three barrier-less reactions. And for KIE our results also agree well compared with the experimental values. For $C + H_2$, both the PES and dynamic calculations are in high accuracy, our Cayley-RPMD gives KIE results within a 4% deviation from experiments. For the other three reaction systems, KIE from Cayley-RPMD are all in high agreement with data from other accurate quantum methods and agree well with experiments. Major deviations would stem from the neglecting of multiple electronic states, which would be our future work. Our results not only provide richer information for the kinetics for all four title reactions, but also show the validity of Cayley-RPMD. Since $C/N/O/S + H_2$ have low barriers or are barrier-less, and all have a deep potential well, it takes a long time to reach convergence so a relatively large amount of CPU time is needed once using standard RPMD. In this work, the obstacle is overcome by Cayley-RPMD, by using a larger time interval of 0.5 fs instead of the original 0.1 fs, without accuracy lost. This work provides another evidence of the capability of Cayley-RPMD in calculations of chemical kinetics.

## SUPPLEMENTARY MATERIAL

See supplementary material for details of table of parameters used in Cayley-RPMDrate calculations; The PMF, transmission coefficient and rate coefficient for X + $H_2$ (400 K) reactions from Cayley-RPMD (0.5 fs) and standard RPMD (0.1 fs), and percentage relative errors; The PMFs and transmission coefficients of Cayley-RPMD for the N + $H_2$ reaction with 0.1 fs, 0.3 fs, 0.5 fs, and percentage relative errors with respect to 0.1 fs; The validation of convergence with the number of beads of both PMF and transmission coefficient curves for X + $H_2$ reaction, and percentage relative errors with 0.5 fs; The PMF and transmission coefficient curves for N + $H_2$ reaction on Ho PES and on Zhou PES; The PMF, transmission coefficient and rate coefficientof N + $H_2$ from Cayley-RPMD and standard RPMD at 400 K on Zhou PES.

## ASSOCIATED CONTENT

**Corresponding Author**

**Yongle Li** – *Department of Physics, International Center of Quantum and Molecular Structures, Shanghai Frontiers Science Center of Quantum, and Shanghai Key Laboratory of High Temperature Superconductors, Shanghai 200444;* orcid.org/0000-0003-0515-6613*; Email:* yongleli@shu.edu.cn

**Authors**

**Wenbin Jiang** – *Department of Physics, International Center of Quantum and Molecular Structures, and Shanghai Frontiers Science Center of Quantum, Shanghai 200444, China*

**Yuhao Chen** – *Department of Physics, International Center of Quantum and Molecular Structures, and Shanghai Frontiers Science Center of Quantum, Shanghai 200444, China*

## ACKNOWLEDGMENTS

This work was supported by the National Nature Science Foundation of China (No. 21973109 and 22173057), the research grant (No. 21JC1402700) from the Science and Technology Commission of Shanghai Municipality. We also gratefully acknowledge HZWTECH for providing computation facilities. The authors also thank helpful discussions from Wenbin Fan at Fudan University and Qiang Li at East China Normal University.